\documentclass[runningheads]{llncs}
\usepackage{graphicx}
\usepackage[linesnumbered,ruled,vlined]{algorithm2e}
\usepackage{algpseudocode}
\usepackage{caption}
\usepackage{subcaption}
\usepackage{amsmath}
\begin{document}
\title{Differential Privacy-based Permissioned Blockchain for Private Data Sharing in Industrial IoT}
\titlerunning{DH-IIoT}
%
\author{Muhammad Islam\inst{1}\and
Mubashir Husain Rehmani\inst{2}\and
Jinjun Chen\inst{3}}
\authorrunning{M. Islam et al.}
%
\institute{Swinburne University of Technology, Hawthorn, VIC 3122, Australia \and
Munster Technological University, Rossa Avenue, Bishopstown, Cork, Ireland \and
Swinburne University of Technology, Hawthorn, VIC 3122, Australia}
%
\maketitle              

\begin{abstract}
Permissioned blockchain such as Hyperledger fabric enables a secure supply chain model in Industrial Internet of Things (IIoT) through multichannel and private data collection mechanisms. Sharing of Industrial data including private data exchange at every stage between supply chain partners helps to improve product quality, enable future forecast, and enhance management activities. However, the existing data sharing and querying mechanism in Hyperledger fabric is not suitable for supply chain environment in IIoT because the queries are evaluated on actual data stored on ledger which consists of sensitive information such as business secrets, and special discounts offered to retailers and individuals. To solve this problem, we propose a differential privacy-based permissioned blockchain using Hyperledger fabric to enable private data sharing in supply chain in IIoT (DH-IIoT). We integrate differential privacy into the chaindcode (smart contract) of Hyperledger fabric to achieve privacy preservation. As a result, the query response consists of perturbed data which protects the sensitive information in the ledger. The proposed work (DH-IIoT) is evaluated by simulating a permissioned blockchain using Hyperledger fabric. We compare our differential privacy integrated chaincode of Hyperledger fabric with the default chaincode setting of Hyperledger fabric for supply chain scenario. The results confirm that the proposed work maintains 96.15\% of accuracy in the shared data while guarantees the protection of sensitive ledger's data.

\end{abstract}
\section{Introduction}
\label{sec:intro}
Blockchain is an emerging technology which has the desirable features of decentralization, tracking, immutability, verification, security, and fault-tolerance \cite{sur4}. Therefore, since its development in financial sector, its adoption in other domains such as IoT, IIoT, e-health, smart grid, and smart city has grown exponentially in the last few years. Due to its attractive and silent features, blockchain is integrated with Industrial Internet of things (IIoT) which results in blockchain-based IIoT. Blockchain-based IIoT enables connectivity of industry partners such as manufacturer, retailer, distributor, and end consumers, to realize a robust supply chain management \cite{blockbasedsupplychaincat1}. However, among the two types of blockchain, i.e., permissioned and permissionless the permissionless blockchain has been criticized for public accessibility. Furthermore, it has several issues such as susceptibility to 51\% attack, low transaction processing rate, and reveal of privacy \cite{sur4,surofconsensusprot}.\\
\indent In supply chain model, different industry partners come together with a shared business interest but have different requirements and policies for privacy preservation of sensitive and secret business data. As a result, certain partners are not willing to share their private and confidential data i.e., business secrets, special offers to certain retailers in a public ledger which is visible to other partners and competitors. For example, in food industry, three separate groups can be considered which are farmer-distributor, distributor-wholesaler, and distributor-retailer.  Companies included in these groups would not be willing to share their business secrets with other participants in the network. In another case, supply chain partners can be competitors and they would not let competitors to see their business plans.\\
\indent To address these issues, permissioned blockchain such as Hyperledger fabric provides two private data sharing mechanisms which are querying mechanism and multichannel mechanism \cite{hyper}. Through querying mechanism, application clients (including third parties) can send queries which are evaluated on the ledger's data. Similarly, in multichannel mechanism for private data sharing between blockchain peers, two levels of privacy and confidential exchange of data exists in Hyperledger fabric which are application-level and data-level. Application-level privacy means a group of participants from network having same business interest come together and establish a private subset of communication which is called channel. Several subsets of private communications are known as multichannel concept in Hyperledger fabric blockchain which results in private communication and data sharing. As a result, communications are limited to valid group members in a channel. A separate and independent ledger is maintained for each such channel. Similarly, for privacy at data-level, Hyperledger fabric uses the concept of private data collection, hash, and transient field in transactions. In this way, only certain participants can see private contents while others can just see hash of the data.\\ 
\indent However, both data sharing mechanisms have serious issues in relation to privacy, utility and transparency of data which need to be addressed. Querying mechanism evaluates query on actual data stored on the ledger through chaincode (smart contract) installed on each peer which makes it inadequate for supply chain environment in IIoT because query response can be utilized by adversaries and suspicious applications to infer sensitive information such as business secrets and special discount offers. Clearly, it needs improvement to make it suitable for supply chain in IIoT. Similarly, multichannel mechanism and private data collection limit utilization of data because the data is confined between restricted parties. As a result, it increases risk of black market and invalid transactions because peers other than restricted group members cannot see the contents of transactions which makes it difficult to verify transactions. In addition, the auditability and accountability are also impacted which are necessary for supply chain such as food industry \cite{sur4,pre-consens}.\\
\indent It is evident from the above discussion that both mechanisms i.e., querying mechanism and multichannel mechanism need further improvement. This is the main motivation of our work in which we target the first mechanism i.e., querying mechanism for improvement in the context of supply chain in IIoT. The contribution of this work is as following: we propose a differential privacy-based permissioned blockchain for private data sharing in the context of supply chain in IIoT by integrating differential privacy into the chaincode of Hyperledger fabric which protects sensitive Industrial data stored on the ledger from linking attacks by adversaries and suspicious applications. We also present an algorithm for accessing the ledger's data in a privacy preserving manner. Similarly, we evaluate the proposed work by implementing a permissioned blockchain using Hyperledger fabric and integrate differential privacy in its chaincode. A privacy threat model based on linking attack is also implemented to evaluate privacy preservation of sensitive data in the context of supply chain in IIoT. Finally, the results are compared with default setting of Hyperledger fabric chaincode (non-privacy preserving) to validate the privacy preservation and utility of data. We prove that the proposed work gets 96.15\% of accuracy in the shared data for $\epsilon$ = 0.5 while guarantees the privacy preservation at the same time.\\  
\indent The rest of the paper is organized according to the following sequence: in Section II, a literature review of the previous works in the related domain is presented. In Section III, we present the proposed work (DH-IIoT) in detail including the working and time complexity analysis of the proposed algorithm whereas Section IV presents evaluation and simulation results. Finally, Section V concludes this work.  
\section{Literature Review}
\label{sec:letr}
In \cite{bcforsmartfac}, a blockchain-based smart factory architecture is proposed with a privacy model to enhance security and privacy. The proposed architecture is lightweight, partially decentralized, easily expandible and have better privacy and security. However, their privacy model is limited to availability, integrity, and confidentiality through encryption, which is an old concept known as confidentiality, integrity, availability (CIA). In \cite{blockbasedsupplychaincat1}, a blockchain-based data sharing scheme in supply chain in IIoT environment is proposed. Attribute-based encryption is adopted to manage the access control of data in IIoT. To automate the flow of goods, smart contract is used in the proposed scheme. However, encryption is used for security of the system and privacy attack model is missing.\\
\indent Similarly, the work in \cite{towardssecureIIoT} has proposed an efficient credit-based proof of work (PoW) consensus for resource constrained environment in IIoT. In addition, a data authority management system is used to protect privacy of sensitive information. The adoptive control of difficulty of puzzle solving in PoW is used which decreases difficulty for honest nodes and increase it for malicious nodes. As a result, the overall efficiency in the system for honest behaviour is enhanced. The throughput of the system is enhanced using directed acyclic graphs instead of conventional blockchain. However, the PoW allows every participant to append a block and validate transactions in which contents of transactions are visible to every node in consensus phase. In addition, encryption is used to protect privacy which is cumbersome to implement.\\
\indent A similar idea has been proposed in \cite{reputationscheme} which works on the principle of rewarding normal behaviour of nodes and punishing abnormal behaviour. The unique feature of their work is that it can be integrated with state-of-the-art PoX protocols. However, privacy concerns associated with PoX protocols are not addressed. A modified bitcoin system known as “Monero” has been proposed in \cite{monero}. Monero ensures privacy and unlikability of transactions to its sender and receiver and protecting balances of participants. However, it uses cryptographic algorithms to hide sending and receiving addresses and balances which are cumbersome to implement. In \cite{materialtrackingcat1}, a smart manufacturing supply chain based on blockchain with traceability and verifiability features has been proposed. The complex and private data sharing is performed by blockchain in a secure manner which enable participants to control data sharing in supply chain. However, the privacy and confidentiality are based on inherent encryption techniques of blockchain and addressing the privacy concern in consensus mechanism is missing. Similarly, the work in \cite{pre-consens} also uses encryption techniques along with Hyperledger fabric to overcome privacy issue during the consensus phase. Finally, Hyperledger fabric adopts hashing and multichannel mechanism to enable private data exchange \cite{hyper}. However, encryption techniques not only significantly impact the utility of data in blockchain but are costly in terms of computation.
\begin{figure}[htp]
\includegraphics[width=\textwidth]{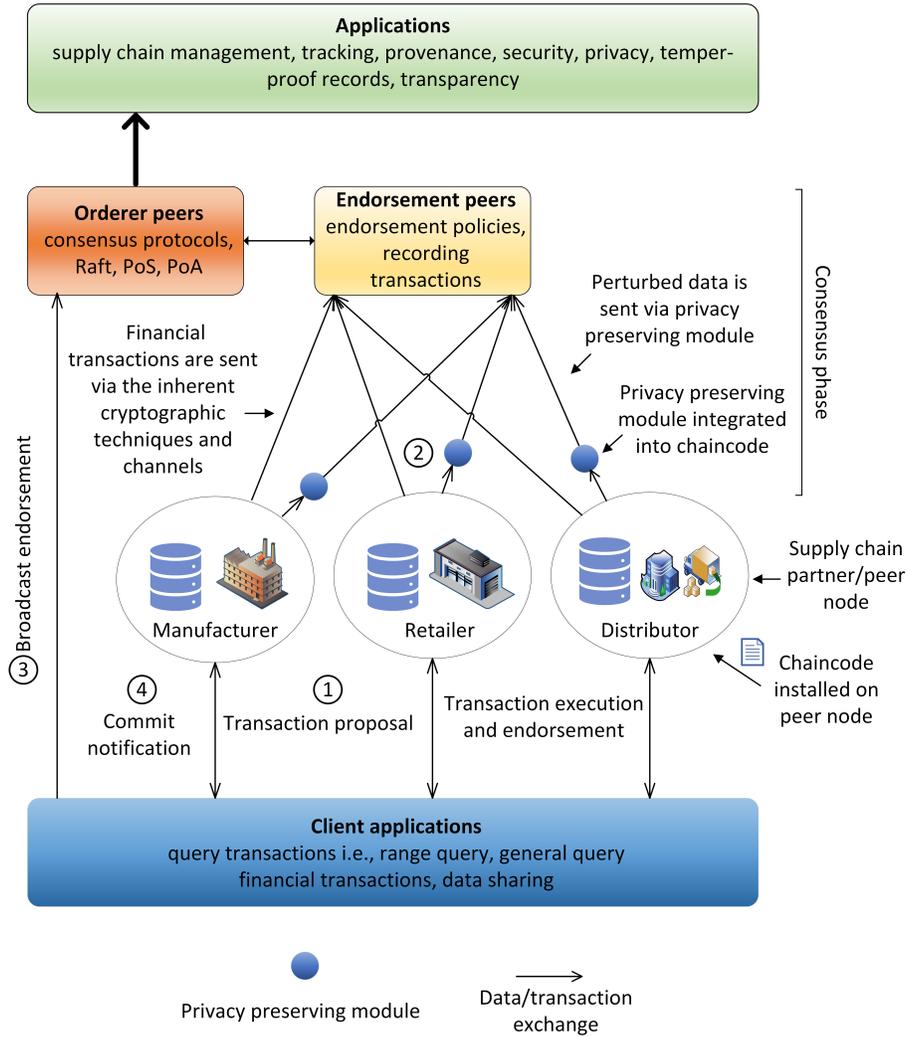}
\caption{System model of transactions flow in privacy preserving chaincode in DH-IIoT.}
\label{fig:sm}
\end{figure} 
\section{Proposed Scheme:DH-IIoT}
\label{sec:pw}
\subsection{System Model}
\label{subsec:sm}
The proposed system model is shown in Fig.~\ref{fig:sm}. The system model is composed of client applications, endorsing peers, privacy preserving module, orderer peers and finally the applications which reside on top of these components. The existing Hyperledger fabric architecture is customized, and a privacy preserving module based on differential privacy is added into the chaincode (smart contract) of Hyperledger fabric. The supply chain participants and third parties act as client applications and a collaboration scenario is also considered in which participants share their data and access data from others through query transactions. Client applications send transaction proposals to peers such as query transactions, range transactions, financial transactions, and data sharing. The transaction proposals are endorsed by the available pool of endorsing peers.\\ 
\indent It is assumed that the peers can differentiate between the mentioned transaction types. As a result, for financial transactions, inherent features of cryptographic techniques and channels are used whereas other transactions are sent via privacy preserving module. In addition, for query transactions only statistical queries are considered in this work, which can be extended for other transaction types as well. An example of a statistical query is how many customers have purchased more than 100 items of a product? The query transaction is represented as $T_{q}=\{f_{1}, f_{2}, f_{3}..f_{n}\}$ where $f_{1}, f_{2}, f_{3}..f_{n}$ represent queries to be evaluated on the ledger data. Similarly, the chaincode instance of the peer evaluates the query on the ledger which is denoted as $R_{q}= \{f^{*}_{1}, f^{*}_{2}, f^{*}_{3}… f^{*}_{n}\}$ where $f^{*}_{1}, f^{*}_{2}, f^{*}_{3}, f^{*}_{n}$ are the answers to queries. This transaction is executed and sent via privacy preserving module of the chaincode before adding it into data block by the orderer peers.
\subsection{Threat Model}
\label{subsec:tht}
In the threat model, the attackers in the proposed scenario launch linking attack. The linking attack is defined as re-identifying individuals or anonymized data from the observed data by combining it with the known knowledge from other sources. In the proposed work, data sharing scenario in the context of supply chain is considered. The supply chain also consists of honest but curious competitors which launch linking attack and try to judge special discounts offered to individuals by other competitors from trade activities using the contents of transactions in data sharing phase. Similarly, third parties which access Industrial data for analysis can also judge personal information and trade secrets from the contents of transactions. We assume a strong attacker that has all other information except the trade activities of target individual. As a result, linking the information with the known data can expose the targeted individuals or trade secrets of other competitors. The proposed DH-IIoT protects data by sharing data through a privacy preserving module based on differential privacy. As a result, attackers cannot expose and isolate a specific individual even all other individuals and trade activities are known to them except their target.
\subsection{Differential Privacy}
\label{dp}
Differential privacy is a popular privacy preservation technique for numerical data. The fundamental principle of differential privacy states that the addition or removal of a single record from a database should not impose significant effect on the final output of a statistical query. According to \cite{dpint2006}, differential privacy can be defined as following:\\
\begin{definition}
``A randomized function Q satisfies $\epsilon$-differential privacy if for all datasets $D_{i}$, $D_{j}$ which differs in one record, and for all $S \subset Range(Q)$, the following holds \cite{dpint2006}:''
\end{definition}
\begin{align}
P[Q(D_{j}) \in S] \leq e^{\epsilon} \times P[(Q(D_{j})\in S] && \text{(according to \cite{dpint2006})}
\label{eq:eqdp}
\end{align}
Where Range (Q) is range of all possible outputs of function Q. $\epsilon$ is known as privacy budget. Moreover, smaller value of the privacy budget is desired to get good privacy. Furthermore, in the literature, two mechanisms have been widely adopted to guarantee differential privacy which are Laplace mechanism and exponential mechanism \cite{differentialpubsurvey}. In this work, we consider the scenario of numerical data sharing, i.e., products rating, number of transactions, count of products, and number of customers.  Therefore, we adopt Laplace mechanism because it is suitable for numerical data perturbation \cite{dwork2014algorithmic}. The Laplace distribution function is given as following:
\begin{equation}
f(x, \mu, \lambda) = \frac{1}{2\lambda}e^{\frac{-|x-\mu|}{\lambda}} \label{eq:lappr}
\end{equation}
Where $\lambda$ is Laplace scale and $\mu$ is mean for Laplace distribution. Furthermore, $\lambda = \frac{{\bigtriangleup}f}{\epsilon}$. In addition, $\bigtriangleup$f is the maximum difference of two queried results from adjacent datasets $D_{i}$ and $D_{j}$. It is called sensitivity and for $D_{i}$ and $D_{j}$ differing in one record it is denoted as following \cite{dpint2006,dwork2014algorithmic}:
\begin{equation}
{\bigtriangleup}f = {| f(D_{i})-f(D_{j}) |}_{1} \label{eq:sensitivity}
\end{equation}
Consequently, the random noise generated from Laplace distribution on scale $\lambda$ is represented as \textit{Lap($\lambda)$} which then added to the actual result.  
\subsection{Working of DH-IIoT}
\label{subsec:wmech}
In Hyperledger fabric, the inclusion of transactions in blocks and validation are performed by separate peers known as orderers and validating or endorsing peers, respectively. This concept enables Hyperledger fabric to get high transaction validation rate and throughput. The working of the proposed DH-IIoT consists of four steps namely (1) transaction proposal, (2) transaction endorsement and privacy preservation, (3) execution of ordering service and (4) transaction validation and commit. In the following, we discuss all these phases along with processing steps and algorithm ~\ref{alg:alg1} for the proposed DH-IIoT in detail.
\subsubsection{Transaction Proposal}
In the first step, application clients send transaction proposal to endorsing peers.  In this phase, ordering service is not involved and it only consists of interaction between application clients and endorsing peers regarding the chaincode function invocation. The set of endorsing peers independently invoke the chaincode with proposal. The set of endorsing peers is chosen according to the endorsement policy defined for the chaincode i.e., one peer from each organization must endorse transaction proposal. Similarly, single endorsing peer can also be targeted which requires only that specific node to endorse the transaction proposal. In proposed scenario, query transaction proposal is considered which is sent to targeted peer or organisation. The ledger state is not altered in this phase because the ordering service is not involved.
\begin{algorithm}
\small
\let\oldnl\nl
\newcommand{\nonl}{\renewcommand{\nl}{\let\nl\oldnl}}
\caption{Differential privacy-based privacy preserving algorithm for DH-IIoT}
\label{alg:alg1}
\SetAlgoLined
\DontPrintSemicolon
\nonl
\textbf{Input:} Ledger data $L_{d}$, query transaction with n queries $T_{q} = \{f_{1}, f_{2}, f_{3}…f_{n}\}$\;
\nonl
\textbf{Output:} Differential private query response $R_{q} = \{f^{*}_{1}, f^{*}_{2}, f^{*}_{3}… f^{*}_{n}\}$\;
\nonl
\textbf{Initialization:} Iteration \textit{i = 1}, random variable \textit{x}, \textit{noise} = 0, mean $\mu$ = 0, Laplace scale $\lambda = \frac{{\bigtriangleup}f}{\epsilon}$, privacy budget $\epsilon$ as given in equation~\ref{eq:lappr}\;
\While {i $\leq$ n}{
\Indp
Execute query $f_{i}$ on the original ledger data $L_{d}$\;
\textbf{Call} LaplacianFunction()\;
Add noise to perturb query response $f_{i} + noise$\;
\textit{i = i + 1}\;
}
\nonl
$\textbf{FUNCTION}\rightarrow LaplacianFunction()$\;  
\Indp
Calculate Laplacian noise using \textit{f(x;$\mu$,$\lambda$)} from equation~\ref{eq:lappr}\;
\KwRet {$noise$}\;
\Indm
\KwRet $R_{q} = \{f^{*}_{1}, f^{*}_{2}, f^{*}_{3}… f^{*}_{n}\}$\;
\end{algorithm}
\subsubsection{Transaction Endorsement and Privacy Preservation}
In our proposed DH-IIoT, the invocation of chaincode with proposal is followed by execution of privacy preserving module as shown in Fig.~\ref{fig:sm}. In this step, the chaincode evaluates query response against the private data stored on the local ledger of the peer. Algorithm~\ref{alg:alg1} which is implemented as a chaincode function is used to add noise to the true answer of query. The query evaluation and noise addition are shown in Fig.~\ref{fig:sc}. Finally, the perturbed response is returned to the requesting client application with endorsement. It is assumed that peer can differentiate between pure financial transaction and a data sharing or query transaction. In this way, financial transactions follow the same existing steps in Hyperledger fabric whereas other query transactions are evaluated using differential privacy module. As a result, every client application can send request to peers for private data access even outside the members list of the channel.  
\subsubsection{Execution of Ordering Services}
In step 3, on receiving the transaction response and enough endorsement from the peers, it is sent to ordering service. The ordering service receives transactions from all channels and combine them in blocks. The ordering service perform sequencing of transactions received from all channels and package them in blocks. Hyperledger fabric gives different options for ordering nodes to carry out consensus on sequencing of transactions i.e., Raft, Kafka, Solo etc. Raft and Kafka both offers fault-tolerance which is beneficial for robust applications across the industry environment. For simplicity reason in proposed work, we adopted single ordering node. In the proposed scenario, the risk of reading the contents of transactions by ordering nodes is avoided by including the perturbed data.
\subsubsection{Transaction Validation and Commit}
In this step, blocks are broadcasted to peers for validation. Each peer validates transactions included in the block and ensures that it meet the endorsement policy. After successful validation, the blocks are committed to peers. The committed blocks are added to the chain which update the status of the ledger. In addition, the blocks which fail the validation phase are not added to the chain. Finally, the application clients are notified of their successful transactions. 
\section{Performance Evaluation}
\label{sec:eval} 
\subsection{Experimental Setup} 
\label{subsec:exs}
The proposed blockchain network consists of two organizations having one peer and a Couch database.  A single channel is maintained between the nodes with one chaincode installed on each peer. The endorsement policy is set to require endorsement of at least one of the two peers. Fabric is used as software under test (SUT) with SDK version 1.4.11. In addition, the Caliper version 0.4.0 is used for evaluation of SUT \cite{caliper}. We used Ubuntu-18 64-bit operating system for our experimental setup. The hardware configuration includes Intel(R)Core (TM) i5-8250U CPU @ 1.6 GHz processor with 8 GB of installed physical memory.
\subsection{Benchmark and Transaction Configuration}
\label{subsec:benchconfig}
The benchmark configuration includes two rounds which are initialization of ledger and querying the ledger. In the first round, a test with five workers is configured to send input transactions with fixed rate in the range of 10-50 tran/sec to the SUT and a total of 500 transactions are sent to initialize the ledger. In the second round, query transaction is configured in which the ledger state is queried by the client application by sending input transactions with fixed rate in the range of 10-50 tran/sec for a total of 15 seconds.\\
\indent It is assumed that the minimum and maximum quantity of product which can be purchased in a single write transaction are 1 and 100, receptively. Similarly, the customer name (owner) in the transactions will be selected from \{Bob, Claire, David, Ali, Alice\} whereas the colours  of the product will be selected from \{red, blue, green, black, white, pink, rainbow\}. In query transaction, the sum of total products purchased by a customer is requested. The differential privacy budget $\epsilon$ is varied in the range 0.5-2.5 to perturb the query response before sending it to the client applications. Moreover, the sensitivity of differential privacy in equation~\ref{eq:sensitivity} is assigned a value of 100 i.e., $\bigtriangleup$f = 100. The reason is that a single transaction removal in the proposed scenario causes a maximum difference of 100 on the sum of total products purchased by a customer.   
\begin{figure}[thp]
\begin{center}
\includegraphics[width=0.70\textwidth]{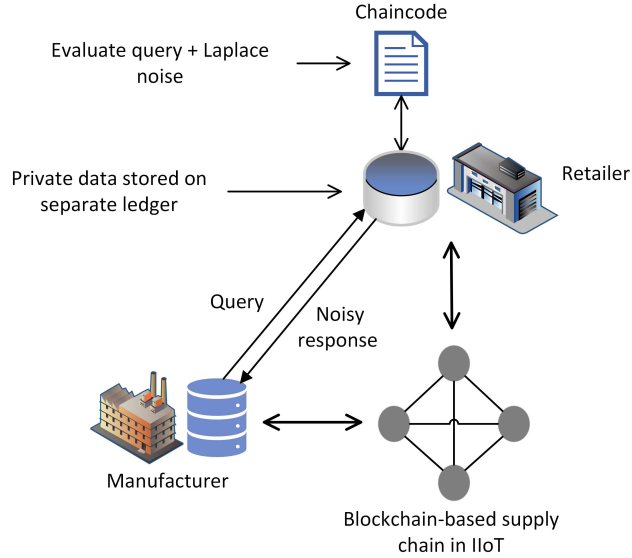}
\caption{Demonstration of query evaluation through chaincode on private data in DH-IIoT.}
\label{fig:sc}
\end{center}
\end{figure}
\subsection{Complexity Analysis}
\label{sec:ca}
In this section, time complexity of algorithm~\ref{alg:alg1} is discussed. The proposed algorithm consists of a single \textit{While} loop which executes according to the number of queries which is denoted as \textit{n} i.e., for each query the loop executes once. Furthermore, each line in the body of the loop takes \textit{O(1)} time to execute. As a result, the time complexity of algorithm~\ref{alg:alg1} is \textit{O(n)}. Therefore, algorithm~\ref{alg:alg1} maintains the high transaction processing rate of Hyperledger fabric. 
\subsection{Simulation Results and Discussion}
\label{subsec:simrsl}
In this section, simulation results obtained from the implementation of the proposed DH-IIoT are presented. The evaluation is performed over four parameters i.e., (1) privacy preservation (2) relative error (3) throughput, and (4) latency of transaction. In the following section, details of the mentioned parameters with comparison results are provided.
\subsubsection{Privacy Preservation}
\label{subsec:pp}
In proposed DH-IIoT, supply chain partners keep separate ledgers for private and public data based on Hyperledger fabric channels and private data collection mechanism. The data stored on public ledger is visible to all other partners however, data stored on private ledger is only visible to restricted members of the channels. For instance, in the group of three supply chain partners namely manufacturer, distributor and retailer, the distributor and retailer maintain a separate private data collection. This data collection is only visible to distributor and retailer. On the other hand, manufacturer can only see the hash of the data \cite{hyper}. In the proposed scenario, manufacturer needs statistical results evaluated on this data to improve its performance. However, both members of the private collection are not welling to share actual data. \\
\indent To access private data of supply chain partners (distributor and retailer in this case), application clients from requesting party (manufacturer in this case) send queries to the peers of associated
\begin{figure}[htp]
\centering
\begin{subfigure}[t]{0.48\textwidth}
\centering
\includegraphics[width=\textwidth]{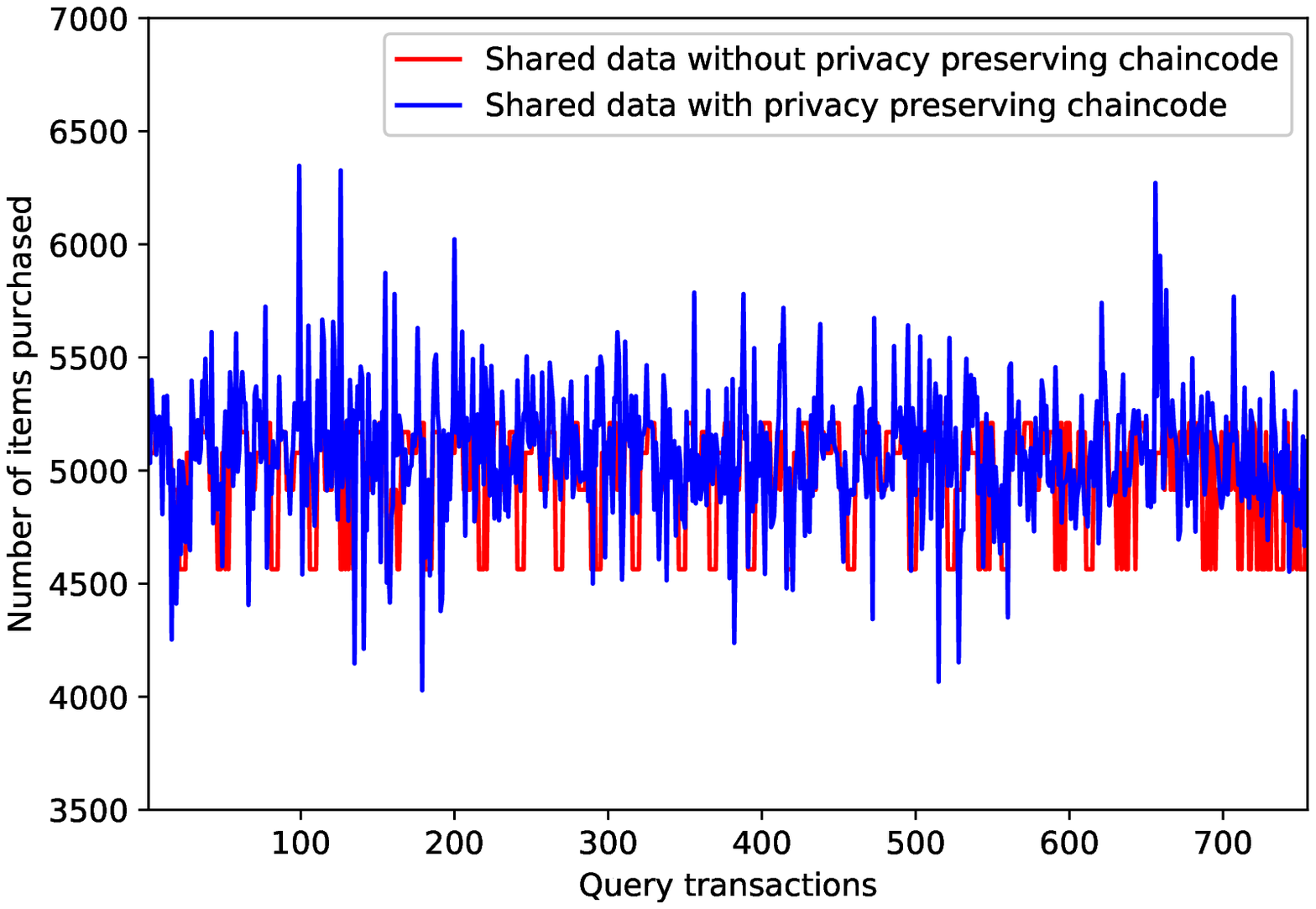}
\caption{protected data for $\epsilon = 0.5$}
\label{dpriv1}
\end{subfigure}
\begin{subfigure}[t]{0.48\textwidth}
\centering
\includegraphics[width=\textwidth]{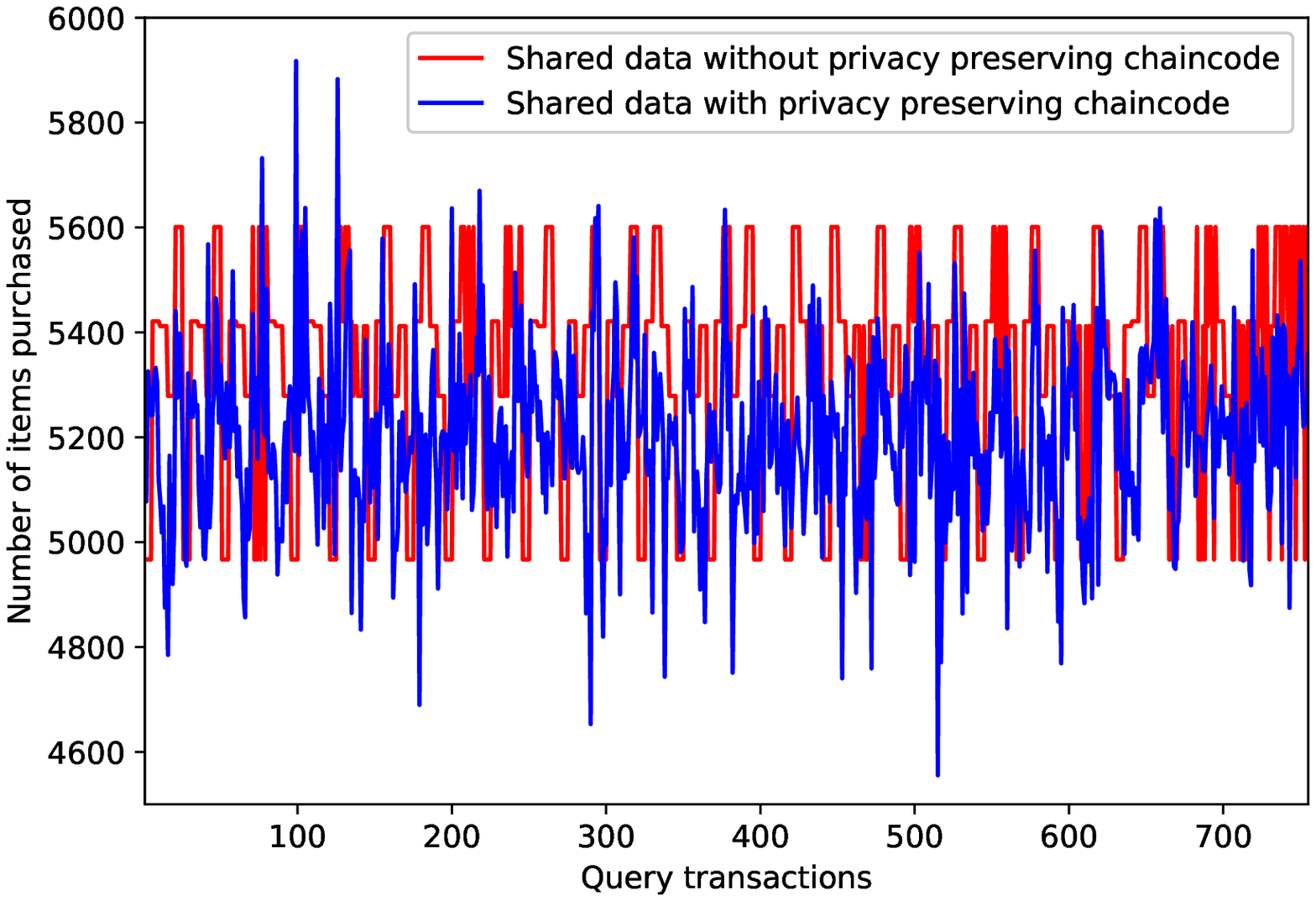}
\caption{protected data for $\epsilon = 1$}
\label{dpriv2}
\end{subfigure}
\caption{Comparison of privacy preserving chaincode (DH-IIoT) with non-privacy preserving chaincode.}
\label{fig:priveval}
\end{figure}
supply chain partners for customers trade activities and number of items purchased. The ledger is populated through
write transactions as discussed in section \ref{subsec:benchconfig}. In this way, the data of trade activities of all customers is maintained in private ledger which results in a dataset with rows and columns. A row represents a write transaction by a customer whereas column is defined according to the transaction fields i.e., product name, customer name, product quantity, and colour. The query transaction requests the sum of quantity i.e., number of items purchased in all transactions by a specific customer. The chaincode executes the query on local ledger of the peer and add random noise generated from Laplace distribution to the true answer using algorithm~\ref{alg:alg1}. The demonstration of query evaluation and noise addition is shown in Fig.~\ref{fig:sc}. The sample noisy responses to query transactions are plotted with varying differential privacy budget $\epsilon$ for default setting of non-privacy preserving chaincode in Hyperledger fabric and privacy preserving chaincode in the proposed DH-IIoT as shown in Fig.~\ref{fig:priveval}.\\ 
\indent It is evident from the comparison results of both chaincodes that increasing the privacy parameter $\epsilon$ decreases the difference between the actual query response and the noisy query response i.e., for $\epsilon$ = 0.5 the variation in query responses is frequent as compared to query responses for $\epsilon$ = 1 as shown in Fig.~\ref{dpriv1} and Fig.~\ref{dpriv2}, respectively. However, the privacy preservation guarantee for $\epsilon$ = 1 is less than $\epsilon$ = 0.5. The reason is that less noise is added for higher values of $\epsilon$. In this way, the adversary is fooled by sending the perturbed query responses in DH-IIoT whereas maintaining almost the same pattern in the shared data. As a result, the adversary will not be able to link it with known data to expose individual's private data i.e., special discount from retailer. Similarly, exposing the individual spending trends or shopping activities are also avoided. However, the addition of noise also impacts the accuracy (utility) of query responses which impacts the utility of data for service and management improvements, future forecast, and quality enhancements through data
\begin{figure}[thp]
\begin{center}
\includegraphics[width=0.80\textwidth]{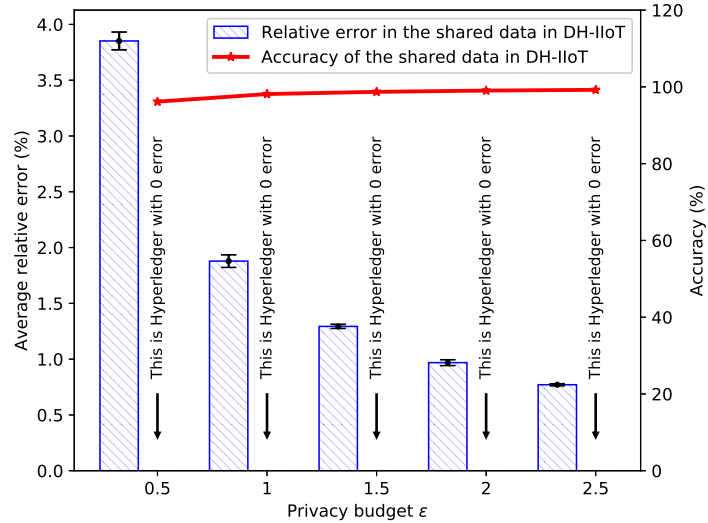} 
\caption{Relative error in query response with varying differential privacy parameter $\epsilon$. The results are within 95\% of confidence interval.}
\label{fig:relerr}
\end{center}
\end{figure}
sharing and data analysis by honest supply chain partners. The more noise added to the data the lower will be the accuracy and vice versa. The trade-off between privacy and accuracy is presented in the next section.
\subsubsection{Relative Error}
\label{subsec:rerr}    
The accuracy of results is measured from the magnitude of relative error in the query response. A high magnitude of relative error means lower accuracy and vice versa. The relative error is defined as following \cite{rerror}:\\
\begin{equation}
\text{Relative error} = \frac{|a - a^{'}|}{a} \times 100\% 
\label{eq:er}
\end{equation}
Where $a$ is the actual value of query response and $a^{'}$ is the perturbed value of query response. Here, 100 is multiplied to get percentage relative error. The average relative error is evaluated with varying $\epsilon$ and the results are shown in Fig.~\ref{fig:relerr}. It is evident from the results that increasing the differential privacy parameter $\epsilon$ decreases the relative error in the shared data through query responses i.e., from 3.85\% to almost 0.75\%.  Similarly, the accuracy is increased from 96\% to 99\%. However, the guarantee of privacy preservation is reduced because less noise is added for higher values of $\epsilon$. Therefore, a trade-off between privacy preservation and accuracy in the shared data exists which should be agreed upon by the data sharing parties. Fig.~\ref{fig:relerr} also shows that more privacy preservation is achieved for sacrificing the accuracy in the query results. As a result, a suitable value of $\epsilon$ should be considered and agreed upon by the data sharing parties.
\begin{figure}[thp]
\centering
\begin{subfigure}[t]{0.48\textwidth}
\centering
\includegraphics[width=\textwidth]{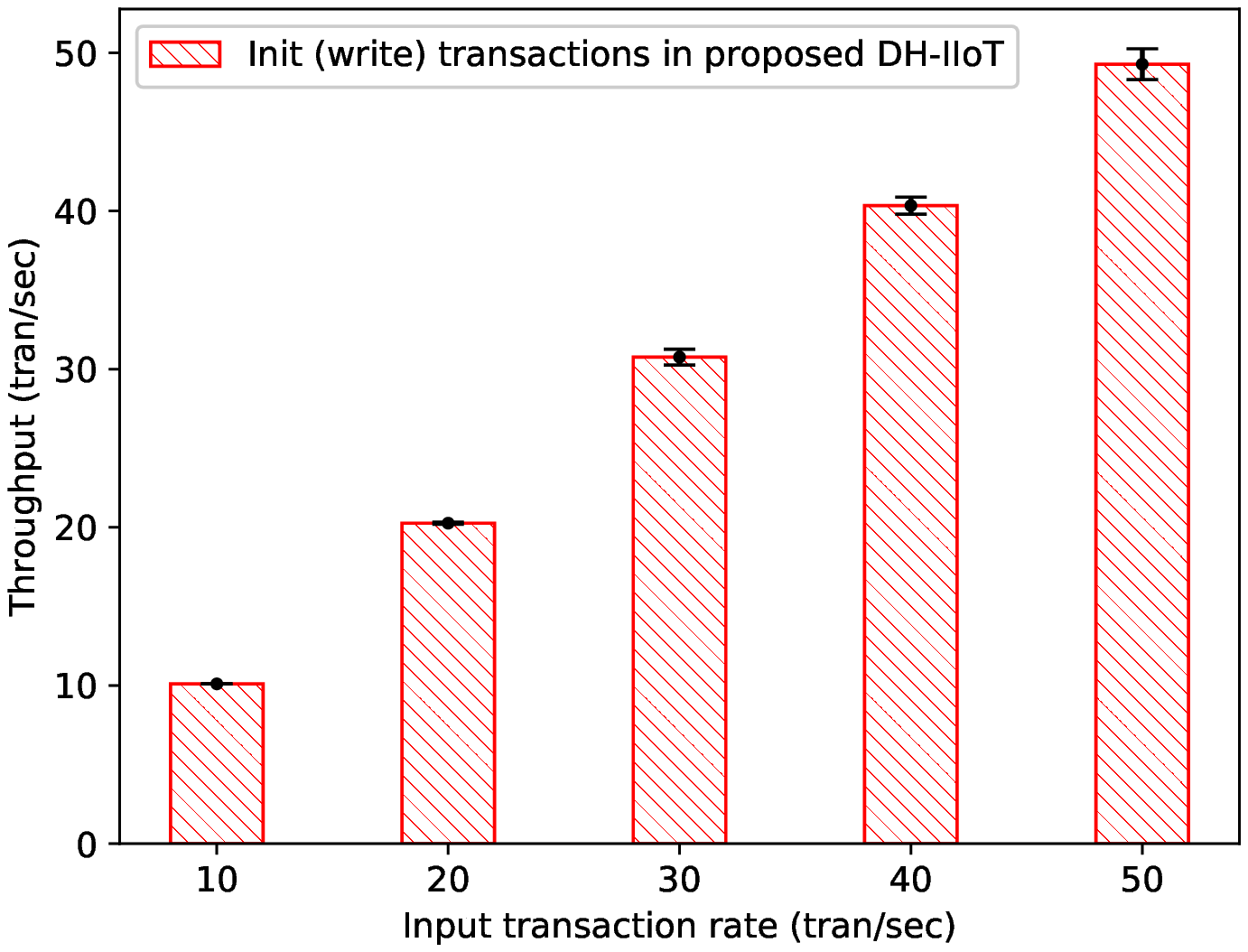}
\caption{}
\label{fig:thpputwrite}
\end{subfigure}
\begin{subfigure}[t]{0.48\textwidth}
\centering
\includegraphics[width=\textwidth]{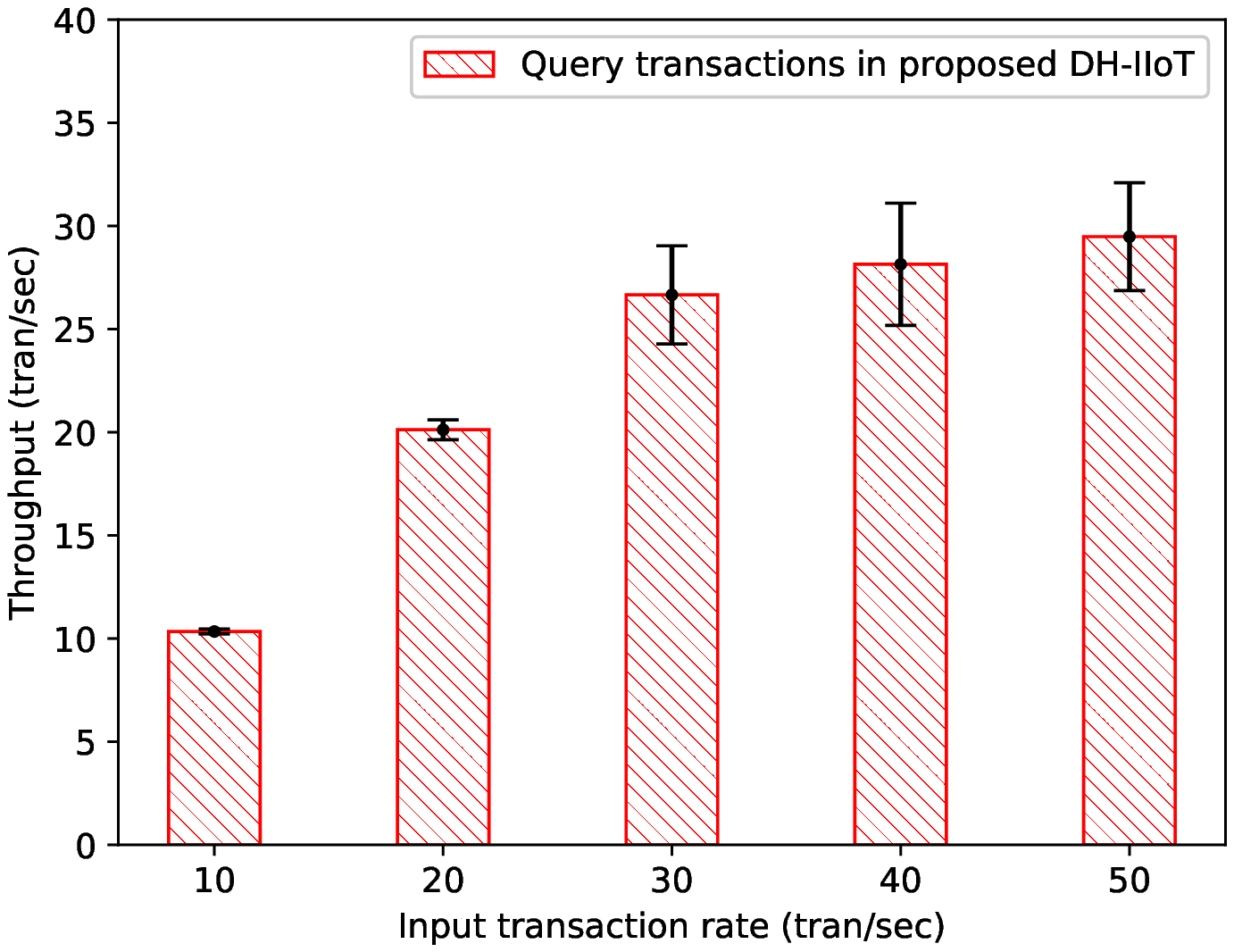}
\caption{}
\label{fig:thputquery}
\end{subfigure}
\caption{Evaluation of throughput in DH-IIoT. The results are within 95\% of confidence interval.}
\label{fig:thput}
\end{figure}   
\subsubsection{Throughput}
\label{subsec:thp}
In this section, throughput of the proposed DH-IIoT is evaluated. In the experimental setup, five workers were configured to send query transactions to the blockchain network (SUT). The chaincode (smart contract) query function is invoked to read the required data from local ledger in a privacy preserving manner using differential privacy module as shown in Fig.~\ref{fig:sc}. The transaction input rate is varied, and the throughput of the blockchain network is evaluated using the Hyperledger Caliper for both Init (write) and query transactions. The results are shown in the Fig.~\ref{fig:thput}. It can be seen from the results that throughput for both Init (write) and query transactions increases with the increase in input transaction rate. The gradual increase in throughput is aligned with the fact that more input transactions in a unit time increase the throughput under the maximum capacity of the network. A maximum throughput of almost 50 tran/sec is achieved in case of write transactions for input transaction rate of 50 tran/sec. Similarly, for query transactions, the maximum throughput is almost 30 tran/ for the same input transaction rate.
\subsubsection{Latency of Transaction} 
\label{subsec:lat}
\begin{figure}[thp]
\centering
\begin{subfigure}[t]{0.48\textwidth}
\centering
\includegraphics[width=\textwidth]{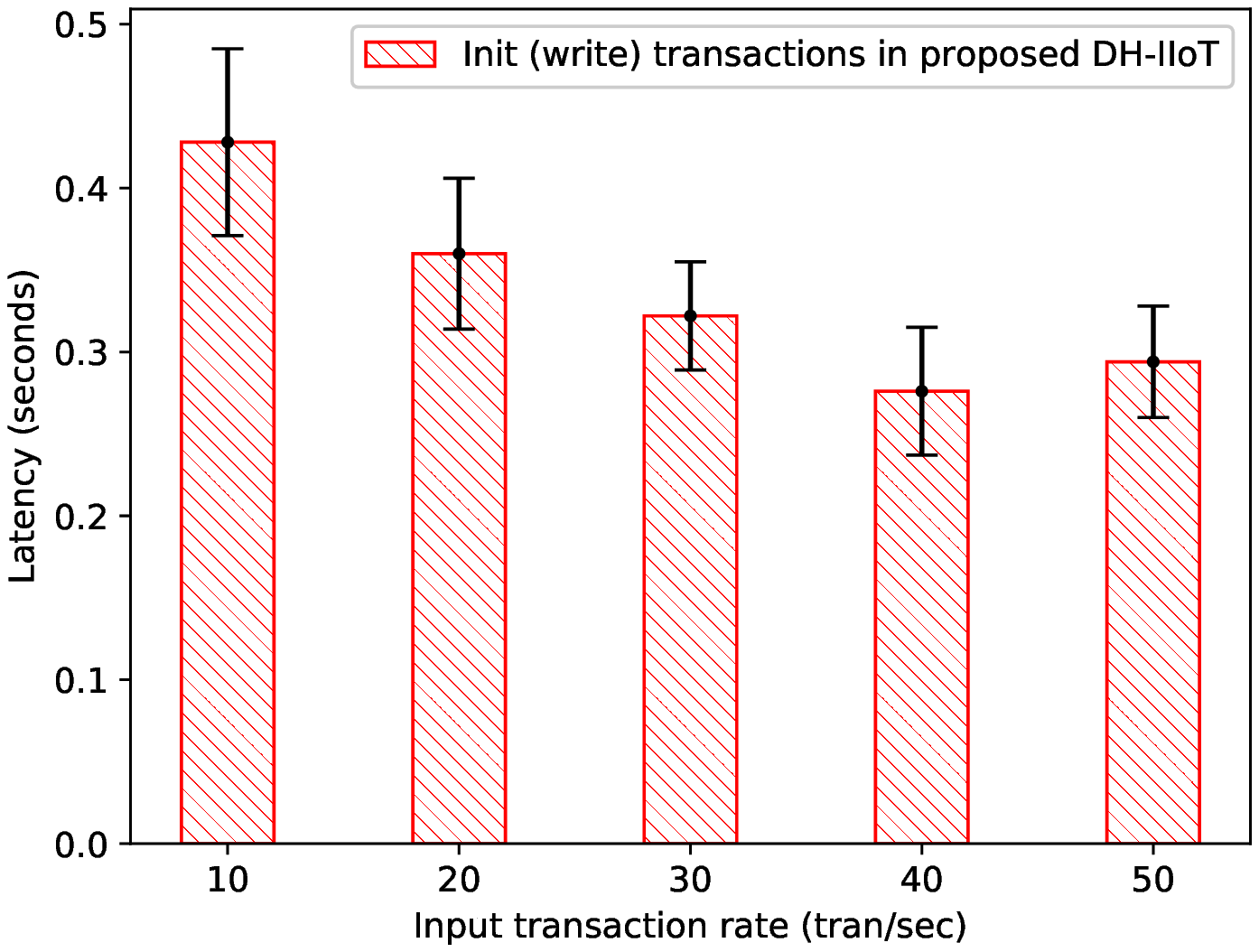}
\caption{}
\label{fig:latencywrite}
\end{subfigure}
\begin{subfigure}[t]{0.48\textwidth}
\centering
\includegraphics[width=\textwidth]{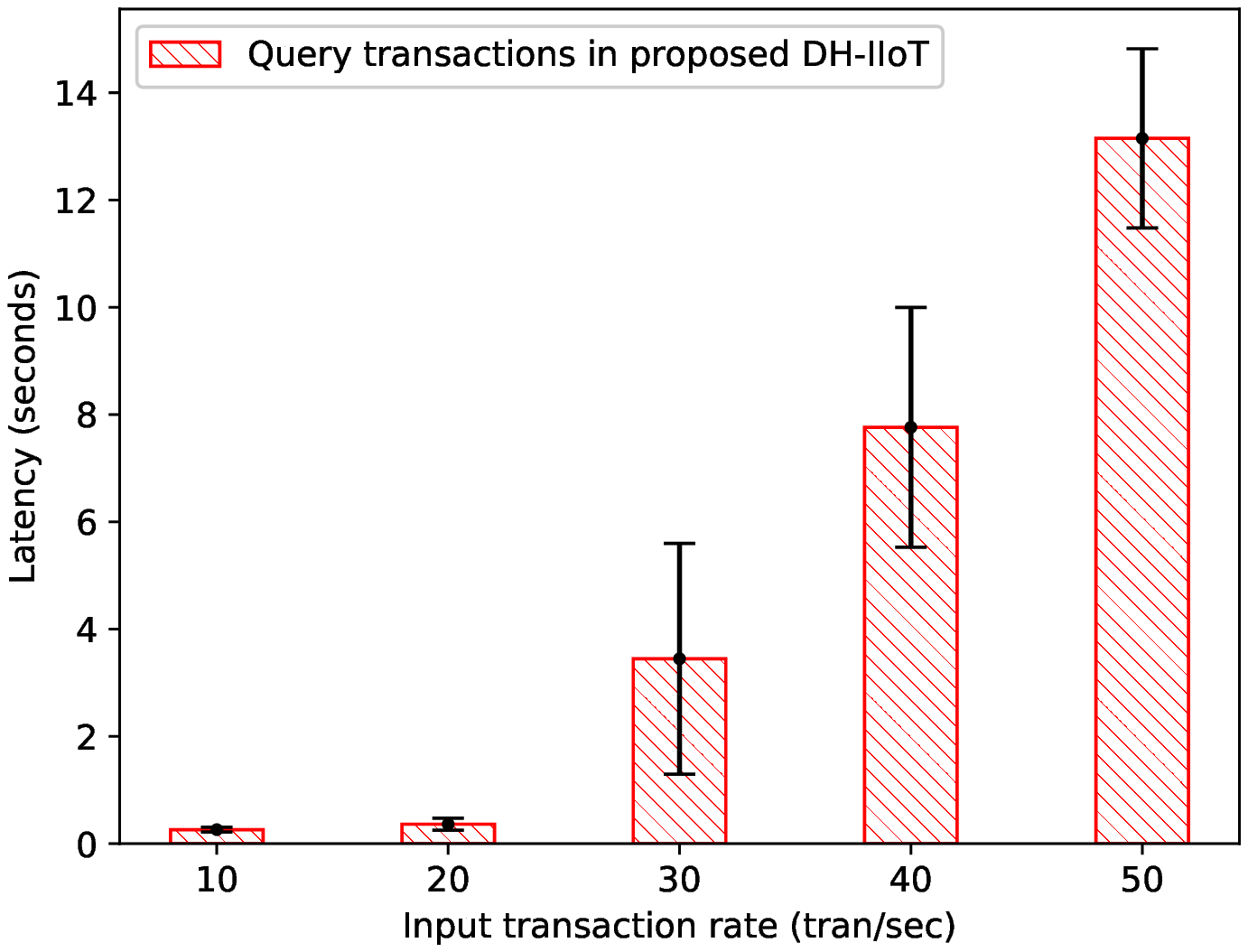}
\caption{}
\label{fig:latencyquery}
\end{subfigure}
\caption{Evaluation of latency of transactions in DH-IIoT. The results are within 95\% of confidence interval.}
\label{fig:latency}
\end{figure}

In this section, latency of transactions in proposed scenario is evaluated. The experimental setting is configured as described in section \ref{subsec:exs}. The latency for both Init(write) and query transactions is evaluated, and the results are shown in Fig.~\ref{fig:latency}. It can be seen from the results that latency for Init (write) transactions
decreases until 40 tran/sec however, it shows an increase beyond this point. The reason is that below the input transaction rate of 40 tran/sec, blockchain network is under the maximum processing capacity and hence transactions take less time for processing. Similarly, the latency of query transactions shows steep increase beyond input
transaction rate of 20 tran/sec. The reason is that on this point, the blockchain network reaches its maximum capacity of processing query transactions and hence beyond this point the transactions latency increases.\\
\indent It is evident from the results of evaluation and comparison that our proposed DH-IIoT enables privacy preservation in a collaborative setting in the context of supply chain in IIoT with inherent features of blockchain such as tracking, validation, querying and recording of transactions. In addition, we improved the performance of existing Hyperledger fabric to enable privacy preservation for marginally compromising the accuracy of shared data. We 
get 96.15\% of accuracy with $\epsilon$ = 0.5 which gives sufficient guarantee of privacy preservation in the shared data. Furthermore, the proposed DH-IIoT enables supply chain partners to record the query response in the ledger which can be used for similar queries sent by other application clients. As a result, it increases the usability of data by re-using the query responses. Finally, DH-IIoT enables supply chain partners and applications to access real-time data recorded on the ledger.             
\section{Conclusion} 
\label{sec:con}
In this work, a differential privacy-based permissioned blockchain using Hyperledger fabric for private data sharing (DH-IIoT) is proposed to solve the issue of exposure of sensitive information in statistical query transactions in the context of supply chain in IIoT. Hyperledger fabric uses two mechanisms for private data sharing which are query mechanism and multichannel mechanism with private data collection for controlling access of confidential data in the network. However, both mechanisms need further enhancement for practical scenarios. In this work, we targeted the querying mechanism of Hyperledger fabric for improvement in the context of supply chain in IIoT. We integrated differential privacy into chaincode of Hyperledger fabric to provide perturbed query responses and protect the original data stored on the ledger. We proved that DH-IIoT maintains 96.15\% accuracy in the shared data for $\epsilon$ = 0.5 which provides sufficient privacy preservation guarantee. The results validated that the proposed DH-IIoT preserves the privacy of sensitive information while maintaining high throughput of the system and improves the performance of Hyperledger fabric.
%
%
%
 \bibliographystyle{splncs04}
 \bibliography{samplepaper.bbl}

\end{document}